\documentclass[aps,pre,reprint,floatfix]{revtex4-2}
\usepackage[utf8]{inputenc}
\usepackage[T1]{fontenc}

\usepackage{amsmath}
\usepackage{amssymb}
\usepackage{amsfonts}
\usepackage{graphicx}
\usepackage{siunitx}
\usepackage{xcolor}
\usepackage{multirow} 
\usepackage{physics} 

\definecolor{darkblue}{rgb}{0, 0, 1}
\definecolor{darkred}{rgb}{1, 0, 0}
\usepackage[colorlinks=true,citecolor=darkblue,urlcolor=darkred]{hyperref}

\newcommand{\K}{K}
\newcommand{\W}{\mathcal{W}}
\newcommand{\B}{\mathcal{B}}

\newcommand{\Q}{\mathcal{Q}}
\newcommand{\LB}{\mathrm{LB}}

\newcommand{\tth}{\textrm{th}}

\begin{document}

\title{Logical and thermodynamical reversibility: optimized experimental implementation of the NOT operation}
\author{Salamb\^{o} Dago}
\author{Ludovic Bellon}
\email{ludovic.bellon@ens-lyon.fr}
\affiliation{Univ Lyon, ENS de Lyon, CNRS, Laboratoire de Physique, F-69342 Lyon, France}

\begin{abstract}
The NOT operation is a reversible transformation acting on a 1-bit logical state, and should be achievable in a physically reversible manner at no energetic cost. We experimentally demonstrate a bit-flip protocol based on the momentum of an underdamped oscillator confined in a double well potential. The protocol is designed to be reversible in the ideal dissipationless case, and the thermodynamic work required is inversely proportional to the quality factor of the system. Our implementation demonstrates an energy dissipation significantly lower than the minimal cost of information processing in logically irreversible operations. It is moreover performed at high speed: a fully equilibrated final state is reached in only half a period of the oscillator. The results are supported by an analytical model that takes into account the presence of irreversibility. The Letter concludes with a discussion of optimization strategies.
\end{abstract}
\maketitle

Starting with a 1-bit information $b\in(0,1)$, only four deterministic information processing operations are possible (Tab.~\ref{1bitop}). Their outcomes are:  initial value $b$ [HOLD], $0$ [RESET to 0], $1$ [RESET to 1] and opposite value $\bar{b}$ [NOT]. The first one is straightforward (do nothing), the second and the third are 1-bit erasures, and the last one corresponds to a bit-flip. It has been shown theoretically~\cite{Landauer_1961} and experimentally~\cite{Berut2012,Berut2015,orl12,Bech2014,Gavrilov_EPL_2016,Finite_time_2020,Hong_nano_2016,mar16,Dago-2021} that erasing a 1-bit memory at temperature $T_0$ requires at least $\W_\LB=k_BT_0 \ln2$ of work, with $k_B$ Boltzmann's constant. This intrinsic and universal minimal energetic cost is known as Landauer's Bound (LB) and comes from the logical irreversibility of a [RESET] operation, or in other words from the entropic loss caused by the reduction of the states available by the system (from two initial states to single reset state). Indeed, the work required to proceed and the heat released during the operation equalize with the entropic loss according to the second law of thermodynamics. On the contrary [HOLD] (or [COPY]~\cite{orl12}) and [NOT] are fully reversible logical operations: they do not come with any information loss. This logical reversibility implies that there is no fundamental minimal bound to the work required to operate. Ref.~\onlinecite{orl12} demonstrated a [COPY] operation with very low cost, below $0.01\, k_BT$. We explore in this Letter the [NOT] operation, that is to say the feasibility of performing a bit-flip in a physically reversible fashion, without spending energy.

\begin{table}[!b]
\caption{The four deterministic 1-bit operations}
\begin{center}
\renewcommand{\arraystretch}{1.1}
\setlength\tabcolsep{1mm}
\begin{tabular}{|c|l|cccc|}
\cline{1-1} \cline{3-6}
Initial &  & \multicolumn{4}{c|}{Final state} \\ \cline{3-6} 
state &  & \multicolumn{1}{c|}{HOLD} & \multicolumn{1}{c|}{RESET to 0} & \multicolumn{1}{c|}{RESET to 1} & NOT \\ \cline{1-1} \cline{3-6} 
0 &  & \multicolumn{1}{c|}{0} & \multicolumn{1}{c|}{0} & \multicolumn{1}{c|}{1} & 1 \\ \cline{1-1} \cline{3-6} 
1 &  & \multicolumn{1}{c|}{1} & \multicolumn{1}{c|}{0} & \multicolumn{1}{c|}{1} & 0 \\ \cline{1-1} \cline{3-6} 
\end{tabular}
\end{center}
\label{1bitop}
\end{table}%

\begin{figure}[t]
	\includegraphics[width=\columnwidth]{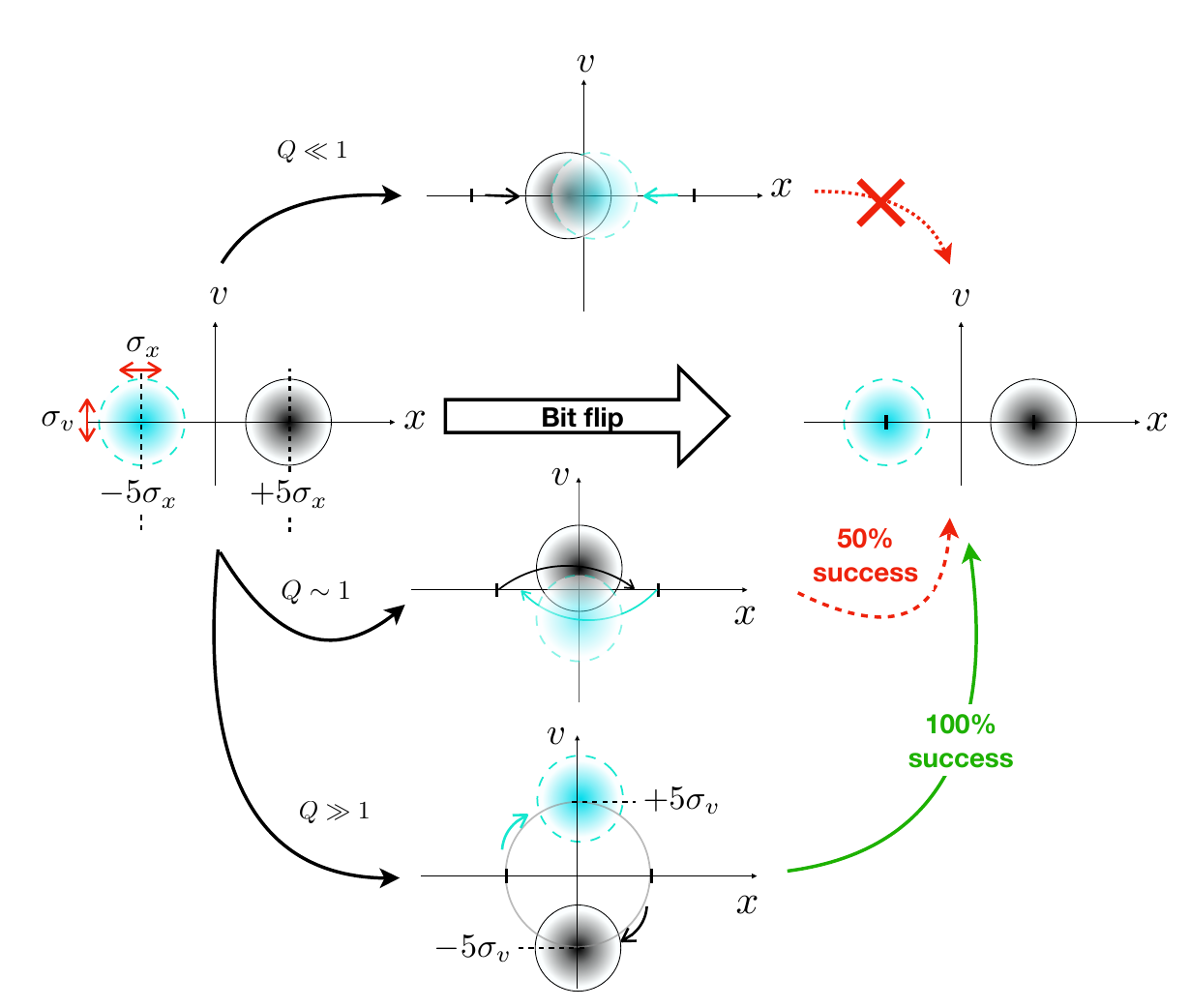}
	\caption{\textbf{Schematic overview of the bit-flip success requirements}. The system two-dimensional PDF are sketched by 2D gaussians in the phase space ($x$,$v$) in light blue, dashed circle (initial state 0) and black, plain circle (initial state 1). Using a single DOF (if $Q\ll1$, only the position can be driven) makes the bit-flip operation impossible: when the system passes through the phase space origin, the Markovian dynamics makes the initial information indistinguishable. The underdamped regime opens a second DOF to process the information: the speed $v=\dot x$. Moderate damping ($Q\sim1$) limits the velocity range accessible and results in a partial overlap of the memory PDF in the two different states: the operation can fail. To prevent the overlap and ensure a $100\%$ success rate, we impose the two states to be separated by a minimal distance: we choose $(\langle x \rangle/\sigma_x)^2+(\langle v \rangle/\sigma_v)^2>25$ as a safety criterion, with $\sigma_x^2$ and $\sigma_v^2$ the position et velocity variance at equilibrium. Bit-flip protocols allowing such velocities require a high enough quality factor of the memory ($Q\gg1$).}
	\label{bitflip_context}
\end{figure}

A bit-flip operation in stochastic but Markovian 1-dimensional (1D) memories, whose dynamics only depends on the current state, is impossible~\cite{bitflip}. Indeed, as sketched in Fig.~\ref{bitflip_context}, a protocol using only one degree of freedom (DOF) has to pass through the same state in the phase space, whatever the initially state (0 or 1):  the information is lost, so the output is random. A second degree of freedom is therefore required to proceed: it can be a second spatial dimension $y$, but if we stick to a 1D memory, observing Non-Markovian dynamics requires the use of the velocity $v=\dot x$. Operating in the underdamped regime (quality factor $Q\gg1$), where the inertia allows the control of the speed (bit-flip conducted in the $x,v$ plan), is therefore a mandatory requirement.

In this Letter, we experimentally implement a bit-flip protocol based on the momentum of an underdamped system, proposed in Ref.~\onlinecite{bitflip} and designed to be reversible at very low damping. We are able to perform a fast and cheap [NOT] operation:  the protocol is performed in the smallest of the typical time scales of the system, and the work required scales as $1/Q$, the prefactor depending only on the memory reliability requirements. Using as 1-bit memory an oscillator of period $\mathcal T_0=\SI{0.68}{ms}$ and quality factor $Q=100$, we perform a bit-flip complying with high standards reliability requirements in only half a period ($\SI{0.34}{ms}$) for an average energetic cost $\langle \W \rangle=\SI{0.46}{k_BT_0}$, significantly below the minimal cost of information processing in logically irreversible operations, $k_BT_0 \ln 2$. The Letter is organized as follows: in a first part, after defining the reliability criteria, we detail the bit-flip protocol designed to be reversible in the ideal dissipationless case. In the second part we implement the protocol experimentally and measure the thermodynamic cost. The experimental results are then supported by a theoretical model that takes into account the presence of irreversibility: the model perfectly matches the experimental results. Finally in the last part we conclude and discuss optimization strategies.

The memory is modeled by a single DOF $x$ evolving in a double well potential $U(x,x_1(t))=\frac{1}{2} k[\lvert x \rvert-x_1(t)]^2$, where $\pm x_1(t)$ set the center of the two quadratic wells controlled by the operator. At rest (before or after the logical operation), we set $x_1 = X_1$ and the potential is $U_1(x)=\frac{1}{2} k(\lvert x \rvert-X_1)^2$. The memory state 0 and 1 correspond respectively to the left and right hand side well, thus to the sign $S(x)$ of $x$. The stiffness of the oscillator $k$ in both wells leads to the position and speed standard deviation at equilibrium: $\sigma_x=\sqrt{k_BT_0/k}$ and $\sigma_v=\omega_0 \sigma_x$, with $\omega_0=2\pi/\mathcal T_0$ the angular resonance frequency in a single well. The reliability of the memory depends on the barrier height $\B=\frac{1}{2}kX_1^2$ between the two states: the higher $\B$, the less probable is a thermal fluctuation high enough to spontaneously flip from one well to the other. We choose $X_1\sim 5\sigma_x$, leading to $\B\sim12.5k_BT_0$: memory losses occur only once every $e^{12.5}\sim\num{3e5}$ relaxation times $\tau_\mathrm{relax}=Q\mathcal{T}_0/\pi$ of the system ($\sim\SI{1.6}{h}$ for our experiment), far beyond any relevant timescale we probe.  

To achieve a high success rate, the bit-flip protocol has to be designed to avoid the overlap of the two possible informations in the phase space as illustrated in Fig.~\ref{bitflip_context}. Indeed a full overlap results in the impossibility mentioned for the single degree of freedom case, while a partial overlap (when the speed is bounded by a moderate damping, $Q\sim1$) decreases the success rate, since the information is likely to slip to the wrong state due to thermal noise. In accordance to the reliability criterion for the static memory, we impose as safety criterion that the center of the two states' probability distribution function (PDF) must be separated at all times by 10 times their characteristic spreading in the phase space (2D gaussian in Fig.~\ref{bitflip_context}), that is to say $10\sigma_x$ and $10\sigma_v$ along the $(x,v)$ axes. It implies that a minimum speed is also imposed to safely convey the information: when $\langle x \rangle =0$, the criterion translates into
\begin{equation} \label{eq:safetyv}
\langle |v| \rangle_{x=0}>5\sigma_v. 
\end{equation}
We therefore need to work in the underdamped regime to allow such high values of the system momentum without having a prohibitive damping cost.

As there is no entropic cost associated to the bit-flip (logical reversibility), the energetic cost of the operation can only come from dissipation during the procedure. There are two strategies to reduce dissipation costs: proceeding at low speed in a quasi-static fashion, or work at very low damping. In 1D, the first strategy has to be eliminated to meet the reliability criterion of Eq.~\eqref{eq:safetyv}. Hence, the only strategy left to maintain physical reversibility without hampering the success rate consists in lowering the viscous damping from the environment: $Q\gg 1$. Within this context, we implement here an innovative bit-flip protocol relying on the system momentum in a non viscous environment to reach physical reversibility while complying with the reliability criterion.

\begin{figure}
	\centering
	\includegraphics[width=\columnwidth]{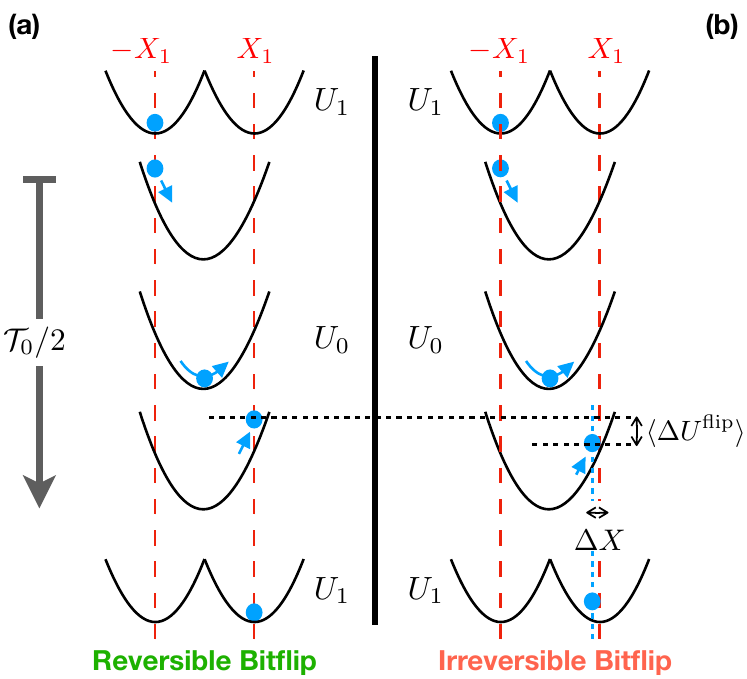}
	\caption{ \textbf{Schematic overview of the bit-flip protocol. (a) Reversible operation (no dissipation)}. The systems starts in state 0 in the encoding potential $U_1$. The operation starts with a sudden change of the potential into a single well $U_0$ centered in $0$. The system without velocity on average therefore initiates an oscillation of period $\mathcal{T}_0$ from the average position $-X_1$. After half a period $\mathcal{T}_0/2$, the trajectory reaches on average  the opposite position $+X_1$ without velocity. At this exact moment, the potential $U_1$ is restored, so that the system ends up at equilibrium in state 1. \textbf{(b) Origin of the irreversibility}. When the system oscillation is damped by the viscous force, the system cannot reach $+X_1$ and culminates at $X_1-\Delta X$. Therefore the operator has to pay for the potential energy difference $\langle \Delta  U^{\textrm{flip}} \rangle$. Besides, the system does not finish in perfect equilibrium in state 1 and has to relax to the well center, only reaching equilibrium in the typical relaxation time $\tau_\mathrm{relax}=Q\mathcal{T}_0/\pi$.}
	\label{schema}
\end{figure}

\begin{figure}
	\centering
	\includegraphics{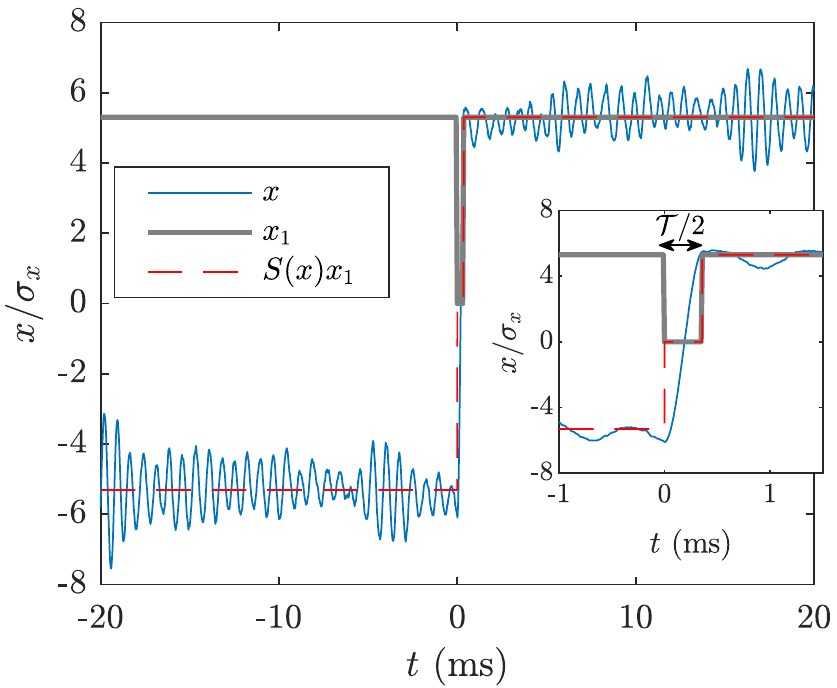}
	\caption{\textbf{Experimental response to the bit-flip protocol}. The bit-flip successfully drives the system from its initial state 0 to state 1 in half a period. The protocol consists in suddenly changing the well center position $x_1$ from $X_1$ to $0$ at $t_i=0$, and changing it back to $X_1$ at $t_f=\mathcal{T}/2=\SI{0.34}{ms}$ (thick grey line). The oscillator trajectory (thin blue line) starts in equilibrium at $\langle x \rangle_i=-X_1$, naturally evolves in the transient single well and ends up at $\langle x \rangle_f =+X_1$. The center of the well in which the cantilever is trapped is plotted in dashed red. The main figure illustrates the initial and final equilibria, while the inset presents a zoom focused on the protocol itself.}
	\label{protocol_bitflip}
\end{figure}

Following Refs.~\onlinecite{bitflip,Ray-2023}, the operation consists in suddenly moving both wells' center to $x_1(t_i)=0$ at the initial time $t_i=0$, as sketched in Fig.~\ref{schema}: the potential becomes a single harmonic well $U_0(x)=\frac{1}{2}kx^2$. After half the oscillator period, at time $t_f=\mathcal{T}_0/2$, the wells' center are brought back to $\pm x_1(t_f) = \pm X_1$ to rebuild $U_1(x)$. We report in Fig.~\ref{protocol_bitflip} the protocol $x_1(t)$ in grey, with an example of a single trajectory in blue, and the corresponding trapping well center ($S(x) x_1(t)$ in dashed red).  It demonstrates the success of the [NOT] operation (here $0 \rightarrow 1$).

This bit-flip protocol, illustrated in Fig.~\ref{schema}, has been designed to be physically reversible when the dissipation can be neglected ($Q \to \infty$). For clarity purposes, let us consider that the system is initially in state 0 (the symmetric case is equivalent): $\langle x \rangle_i=-X_1$ and $\langle v \rangle_i=0$. At $t_i$, the center of the well is suddenly changed form $x_1=-X_1$ to $x_1=0$, the cantilever starts an oscillation into the single well potential. After half a period it reaches in average the opposite maximal position without speed: $\langle x \rangle_f=+X_1$, $\langle v \rangle_f=0$. The second change of the potential at this exact moment therefore doesn't affect the average position of the system, nor its velocity: the memory is immediately in equilibrium. Let us point out that between the two changes, the velocity reaches $\langle \lvert v \rvert \rangle = 5 \sigma_{v}$ when $\langle x \rangle =0$ as required by the safety criterion recalled in Fig.~\ref{bitflip_context}. The operation results in changing the position of the oscillator from $\langle x \rangle_i=-X_1$ to $\langle x \rangle_f=+X_1$ using only the free evolution of the system inside the potential $U_0$: it is a reversible bitflip. 

In $U_1$ the second well is statistically inaccessible, hence the potential remains in practice quadratic with a constant stiffness during the operation. The Fokker-Planck equation ruling the stochastic dynamic is thus linear: the system response is at all time the sum of the deterministic contribution $x_D=\langle x \rangle$ and of the thermal stochastic one $x_{\tth}$: $x=x_D+x_{\tth}$. The latter is not impacted by the bit-flip protocol and remains at equilibrium: $\langle x_{\tth}^2 \rangle =\sigma_x^2=k_BT_0/k$. Therefore the dynamics is ruled by the deterministic trajectory of the oscillator.

In the ideal case without any dissipation, the energy given to the system at the first potential change is fully recovered when $U_1$ is restored: the operation  is reversible and no work is required for the process. Formally as the changes are instantaneous, the work corresponds to the potential loss $\Delta  U^{\textrm{flip}}$ during the flip: 
\begin{subequations} \label{WandUflip}
\begin{align}
\W &= U_0(t_i) - U_1(t_i)+ U_1(t_f) - U_0(t_f)\\
\langle \W \rangle &=-\langle \Delta  U^{\textrm{flip}} \rangle
\end{align}
\end{subequations}
Since $\langle U (t) \rangle = \frac{1}{2}k \left[(x_D(t) - x_1(t))^2 +\langle x_{\tth}^2\rangle\right]$, without dissipation we have:
\begin{align}
\langle U_1 (t_i) \rangle = \langle U_1 (t_f) \rangle_{Q=\infty} &= \frac{1}{2}k_BT_0,\\
\langle U_0 (t_i) \rangle = \langle U_0 (t_f) \rangle_{Q=\infty} &= \frac{1}{2}(k X_1^2+k_BT_0),
\end{align}
so that $\langle \W \rangle_{Q=\infty} = -\langle \Delta  U^{\textrm{flip}} \rangle_{Q=\infty} = 0$.

\begin{figure}[t]
	\centering
	\includegraphics[width=\columnwidth]{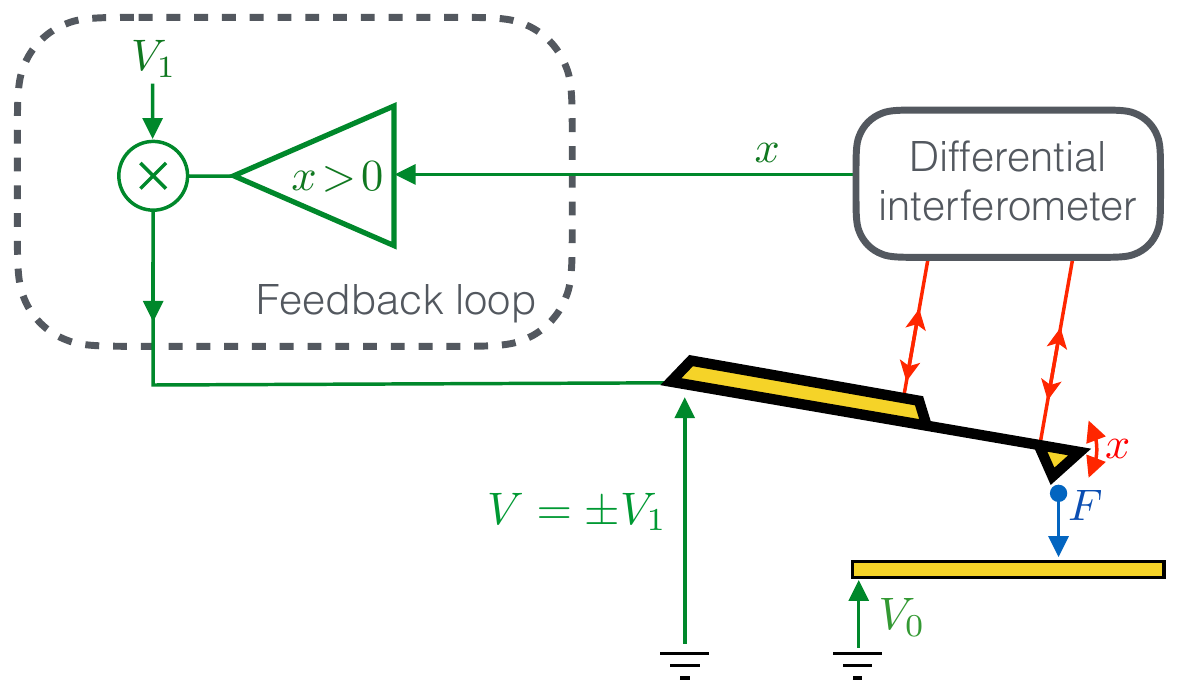} 
	\caption{\textbf{Experimental setup}. The deflexion $x$ of a conductive cantilever is measured with high precision by a differential interferometer. A voltage $V_0\pm V_1$ (with $V_0 \gg V_1$) is applied between the cantilever and a facing electrode by a fast feedback loop, it centers the oscillator around $\pm X_1$ according to the sign of $x$~\cite{Dago-2022-JStat,Dago-2023-AdiabaticComputing}.. The protocol is performed by turning off the feedback (setting $V_1=0$) during half an oscillation period.}
	\label{Fig:setup}
\end{figure}

In our experiment, the 1-bit information is encoded into the position $x$ of an underdamped micro-mechanical oscillator of effective mass $m$, in the form of a micrometric cantilever~\cite{Dago-2021,Dago-2022-JStat,Dago-2022-PRL,Dago-2023-AdiabaticComputing}. The natural angular resonance frequency of the oscillator is $\omega_0=2\pi\times\SI{1.39}{kHz}$, and its low stiffness $k=m\omega_0^2$ results in $\sigma_x \lesssim \SI{1}{nm}$ at room temperature $T_0=\SI{295}{K}$. The quality factor is tuned to $Q=100\pm5$ using a low vacuum environment (pressure $\SI{1}{mbar}$). The deflection $x$ is precisely measured by interferometry~\cite{Paolino2013}, and the double well potential is created by applying an electrostatic force driven by a fast feedback loop based on the comparison of $x$ with 0: $x>0$ (resp. $x<0$) results in a constant force centering the well in $+x_1$ (resp. $-x_1$). The setup is sketched in Fig.~\ref{Fig:setup} and described in greater details in Refs.~\onlinecite{Dago-2022-JStat,Dago-2023-AdiabaticComputing}.

\begin{figure}[t]
	\centering
	\includegraphics{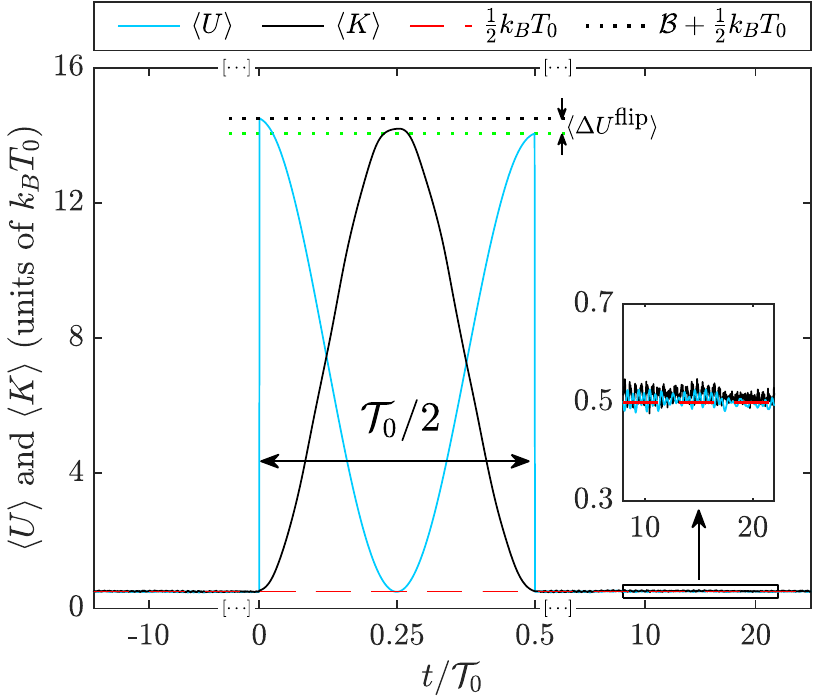} 
	\caption{\textbf{Average potential and kinetic energies}. $\langle U \rangle$ (light blue) and $\langle K \rangle$ (black) are averaged on $N=2000$ trajectories. The bit-flip protocol takes place between $t_i=\SI{0}{ms}$ and $t_f=\mathcal{T}_0/2=\SI{0.34}{ms}$. Before and after the protocol (note the non linear scale for these time intervals, displaying a long time trace, with a vertical zoom in the inset), the oscillator is at equilibrium and equipartition applies: $\langle U \rangle= \langle K \rangle=\frac{1}{2}k_BT_0$ (red dashed). During the protocol, the potential energy gains the barrier height $\B=14k_BT_0$ at $t_i=\SI{0}{ms}$ when the potential is changed from $U_1$ to $U_0$. The oscillator then evolves in an harmonic well for half a period, and energies display the deterministic evolution $U_D=\frac{1}{2}kx_D^2$ and $K_D=\frac{1}{2}mv_D^2$: the potential energy reaches its minimum when the kinetic is maximum, and increases again till the second potential peak (light green dotted line) as the system reaches the opposite position. At $t_f$, the average speed is $0$, thus the kinetic energy recovers its average equilibrium value $\frac{1}{2}k_BT_0$ (red dashed). The potential energy doesn't perfectly recover its initial value, as the small damping removes an energy $\langle \Delta U^{\textrm{flip}}_{\textrm{meas}} \rangle=(-0.450\pm0.002)k_BT_0$ between the two extreme potential values.}
	\label{U_and_K_bitflip}
\end{figure}

The initial distance is calibrated using the position signal during the equilibrium steps: $X_1=\SI{5.3}{\sigma_x}$, so that $\mathcal{B}=(14.00\pm0.05)k_BT_0$. We record $N=2000$ trajectories, alternating between $0 \rightarrow 1$ and $1 \rightarrow 0$ operations. The protocol success rate is $100\%$: none of the 2000 trajectories ended in the wrong final state. We use the experimental data to compute the average potential and kinetic energies $\langle U \rangle$ and $\langle K \rangle =\frac{1}{2}m \langle v^2 \rangle$ displayed in Fig.~\ref{U_and_K_bitflip}. Both quantities present a one cycle oscillation of amplitude $\B$ during the protocol from $t_i=0$ to $t_f=\mathcal{T}_0/2$, and immediately go back to their equilibrium value $\frac{1}{2}k_BT_0$ prescribed by the equipartition at $t_f$. 

We compute the work and heat in the stochastic thermodynamic framework~\cite{sek10,Dago-2021} and extract their distributions on the $N$ trajectories. In Appendix~\ref{appendix:WandQ} we plot these distributions and study the contributions of the intrinsic thermal noise and the extrinsic measurement noise. We measure:
\begin{subequations} \label{WandQmeas}
\begin{align}
\langle \W \rangle &=(0.46 \pm 0.04) k_BT_0\\
\langle \Q \rangle &=(0.43 \pm 0.04) k_BT_0.
\end{align}
\end{subequations}
In the data analysis, we use measured quantities for all variables, including $x_1(t)$: we take into account the finite speed $\dot{x_1}$ when switching back and forth between $X_1$ and $0$. Let us point out that our work measurement is independent of the potential energy computation of Eq.~\eqref{WandUflip}. Note that the average stochastic heat is computed on a time window spanning several relaxation times after the end of the protocol, to allow the memory to return to equilibrium. Shorter integration times (limited to $t_f$ for example) would bias the result, as heat exchanges are slow for underdamped systems~\cite{Dago-2022-PRL}. As the system initial and final states are at equilibrium at temperature $T$ in the same potential $U_1$, the initial and final potential and kinetic energies are equal:  $\langle\Delta U\rangle = \langle\Delta K \rangle= 0$. The first law of thermodynamics thus impies that the average heat dissipated at the end of the procedure is equal to the average work required : $\langle \Q \rangle=\langle \W \rangle$, as measured experimentally. These non null values can be explained by the small residual damping at the origin of the irreversibility, as detailed in the following section. Nevertheless our experimental implementation of the bit-flip already requires less energy than the landmark cost for irreversible operations on a 1-bit memory, Landauer's bound $\W_\LB\sim 0.69 k_B T_0$. Furthermore, this logically reversible operation is performed in a very short time ($\SI{0.34}{ms}$ here). Carrying out an irreversible operation such as an erasure on a similar duration would lead to exceed LB by several $k_BT_0$~\cite{Dago-2021,Dago-2022-PRL}. Finally, the equilibrium is restored just after the procedure, so that bit-flips can be repeated successively without altering the memory reliability.

We tackle in this paragraph the origin of the irreversibility detected through the work and heat mean values: it is the residual damping in the vacuum in which evolves the cantilever. Indeed because of the dissipation during the half oscillation, the potential energy given back by the system is lower than the one initially given by the operator, so that in total, work is required to proceed. To phrase it differently: the damped oscillator launched in $-X_1$ stops at zero speed after half a period a little bit before the exact opposite position $+X_1$ as sketched on Fig.~\ref{schema}(b). 
To provide a quantitive description, let us express the deterministic term of trajectory $x_D$ during a $0\rightarrow1$ operation. The oscillation initiated in $x_D(0)= -X_1$ and $v_D(0) =0$ obeys :
\begin{align}
x_D(t)=X_1e^{\frac{-t\omega_0}{2Q}}(\frac{\omega_0}{2Q\Omega} \sin{\Omega t} - \cos{\Omega t})
 \label{xDbitflip}
\end{align}
where $\Omega =\omega_0 \sqrt{1-1/(4Q^2)}$ is approximately $\omega_0$ at high quality factor. After half a period, at $t_f = \mathcal{T}/2 = \pi/\Omega$ the cantilever reaches on average the extreme position:
\begin{align}
\langle x (t_f) \rangle= x_D(\mathcal{T}/2)=X_1e^{-\pi/\sqrt{4Q^2-1}}.
 \label{xDT0}
\end{align}
Without damping, $Q\rightarrow \infty$ so that we recover $\langle x \rangle_f=X_1$ and consequently a reversible behavior. Meanwhile in a viscous environment, the cantilever undershoots the targeted position by $\Delta X = X_1(1-e^{-\pi/\sqrt{4Q^2-1}})\simeq \frac{\pi}{2Q} X_1$. For $Q=100$, we have $\Delta X/X_1=1.56\%$. As a consequence, there is a potential energy loss $\langle \Delta  U^{\textrm{flip}} \rangle$ that we compute using Eq.~\eqref{WandUflip} as:
\begin{align}
\langle \Delta  U^{\textrm{flip}} \rangle&=-\frac{1}{2}k \left[X_1^2 + (\langle x (t_f) \rangle-X_1)^2 - \langle x (t_f) \rangle^2\right] \\
&=-k X_1^2\left(1-e^{-\pi/\sqrt{4Q^2-1}}\right)\simeq-\frac{\pi}{Q} \mathcal{B} \label{DUflip}\\
&=(-0.44 \pm 0.02) k_B T_0, \label{DUflip_value}
\end{align}
where the approximation in Eq.~\eqref{DUflip} is true for $Q\gg1$. The value in Eq.~\eqref{DUflip_value} corresponds to the theoretical prediction knowing the parameters $Q$ and $\mathcal{B}$ from calibration, and matches the measured stochastic work and heat reported in Eq.~\eqref{WandQmeas}. We also compare it with the experimental value measured of the experimental potential energy evolution displayed on Fig.~\ref{U_and_K_bitflip}:
\begin{equation} \label{DUmeas}
\langle \Delta U^{\textrm{flip}}_{\textrm{meas}} \rangle=(-0.450\pm0.002) k_BT_0
\end{equation}
The errors are inferred from the error on $\sigma_x$ calibrated before each of the $N$ operations. The theory and the experiment are in very good agreement. Besides, the first peak value in Fig.~\ref{U_and_K_bitflip} is also consistent with the model, being worth the thermal energy plus the barrier energy (deterministic contribution): $\mathcal{B}+\frac{1}{2}k_BT_0=\SI{14.5}{k_BT_0}$.

\begin{figure}
	\centering
	\includegraphics{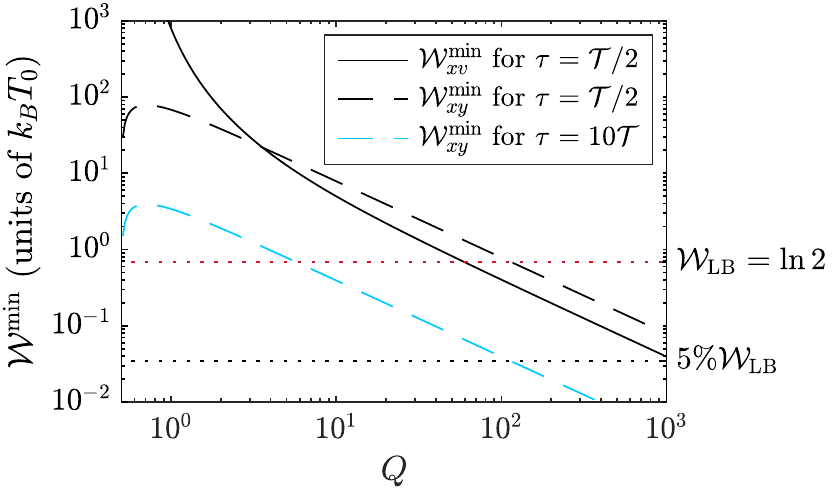}
	\caption{\textbf{Work of a Bit-flip protocol depending on the quality factor $Q$ and the operation duration $\tau$}. Our protocol (plain black) allows very fast erasures: $\tau=\mathcal{T}/2$. It corresponds to a bit-flip in the $(x,v)$ plane, whose cost to ensure the safety criterion is given by Eq.~\eqref{xvplan}. Increasing the quality factor reduces the bit-flip cost: the quasi-reversibility  ($\langle \W \rangle<5\%  \W_\textrm{LB}$ in black, lower dotted line) is reached for $Q>1000$. Proceeding at the same speed using the 2D spatial alternative (bit-flip in the $(x,y)$ plane, dashed black) requires twice more energy at large $Q$, as expressed in Eq.~\eqref{xyplan}. On the contrary at small $Q$ or lower speed (for example for $\tau=10 \mathcal{T}$, dashed light blue) the $(x,y)$ bit-flip is cheaper than the momentum based protocol.}
	\label{bitflipopti}
\end{figure}

It can be noted when $Q$ is not very large, the undershoot $\Delta X$ can be significant: it diverges for $Q=\frac{1}{2}$, when the motion is not underdamped anymore. Moreover, the protocol takes longer, as the effective period $\mathcal{T}$ increases when $Q$ decreases. At low $Q$ values, meeting the safety criteria at $t_f$ implies choosing $X_1\geqslant 5\sigma_x e^{\pi/\sqrt{4Q^2-1}}$ to compensate for the decrease in amplitude. From Eq.~\eqref{DUflip}, we compute the minimum value for the mean work:
\begin{equation} \label{xvplan}
\W_{xv}^\mathrm{min} = 25 k_B T_0 e^{2\pi/\sqrt{4Q^2-1}} \left(1-e^{-\pi/\sqrt{4Q^2-1}}\right).
\end{equation}
In agreement with Eq.~\eqref{xvplan}, the best way to cut the bit-flip cost is to enhance the quality factor as displayed in Fig.~\ref{bitflipopti}. In particular, to ensure less than $5\%$ of the LB, the quality factor has to exceed $Q=1000$. 

As mentioned in the introduction, using a second spatial DOF~\footnote{For a colloidal particle optically trapped, the $y$ coordinate is a natural second DOF. For a micro-beam oscillator, this could also be the case for a rod (instead of a cantilever), with equivalent stiffnesses in the 2 directions perpendicular to its length.} is an alternative to the speed DOF. Maintaining the bit-flip success rate ensured by the safety criterion ($2X_1=10 \sigma_x$ between the two states at all time) would in this case cost at least the work required to proceed along a circle in the $(x,y)$ 2D plane, thus covering a distance $\pi X_1$ in a time $\tau$. Bypassing transients using optimized protocols~\cite{Gomez} and moving at constant speed, the work implied is
\begin{align}
 \W_{xy}^\mathrm{min} & = k \frac{(\pi 5\sigma_x)^2}{Q \tau \omega_0} =25 k_B T_0  \frac{\mathcal{T}_0}{\tau} \frac{\pi}{2Q}. \label{xyplan}
\end{align}
As illustrated on Fig.~\ref{bitflipopti}, for operations as fast as $\tau=\mathcal{T}/2$, the protocol based on momentum is better than its spatial $(x,y)$ counterpart, as long as $Q>3.6$. However, the bit-flip in the $(x,y)$ plane allows one to reduce the operation speed and get close to a quasi-static motion of the system in the viscous bath. As shown in Fig.~\ref{bitflipopti}, if one accepts to extend the duration to $\tau=10 \mathcal{T}$, then the $(x,y)$ bit-flip protocol is better. In particular, such a slow process would reached the quasi-reversibility ($\langle \W \rangle<5\% \textrm{LB}$) in our experimental conditions, \textit{i.e.} $Q=100$.

To summarize, we have experimentally illustrated the connection between physical reversibility and logical reversibility in information processing. The bit-flip protocol designed to be secure and cost-free~\cite{bitflip,Ray-2023} has been tested experimentally: it successfully performs the NOT operation in $\mathcal{T}_0/2=\SI{0.34}{ms}$ for a very small amount of work, for example significantly below the landmark of LB of irreversible operations. The deviation from the desired zero-work operation is fully explained by the coupling of the memory to the surrounding bath: even very low damping introduces irreversibility. The theoretical description has proven reliable to quantify the remaining irreversibility, evaluated with high accuracy in our experiment by four independent measurements, all in agreement [Eqs.~\eqref{WandQmeas}, \eqref{DUflip_value} and \eqref{DUmeas}]. Even faster and cheaper NOT operations could be achieved with oscillators having a higher resonance frequency and a larger quality factor. After the reset operation~\cite{Dago-2021,Dago-2022-PRL,Dago-2023-AdiabaticComputing}, this Letter demonstrates the last logical operation on single bit underdamped memories, and further highlights their low energy footprint and interest.

\vspace{0.2cm}

\acknowledgments

This work has been financially supported by the Agence Nationale de la Recherche through grant ANR-18-CE30-0013. We thank C. Plata for suggesting the study of reversible operations and providing relevant references, J. Pereda for the initial programming of the digital feedback loop creating the virtual potential, and S. Ciliberto for fruitful scientific discussions and useful remarks on the manuscript.

\appendix

\section*{APPENDIX}

In this appendix, we study the probability distribution function of the work during a bitflip protocol. We show that the measurement error, even small, proscribes the use of Jarzynski's equality. Indeed, since the width of the distribution is expected to scale as the inverse of the quality factor of the oscillator, for underdamped systems the measurement noise plays a significant role and bias the statistics of the exponential averaging needed to apply Jarzynski's equality.

\section{Experimental work and heat distributions} \label{appendix:WandQ}

\begin{figure}
	\centering
	\includegraphics{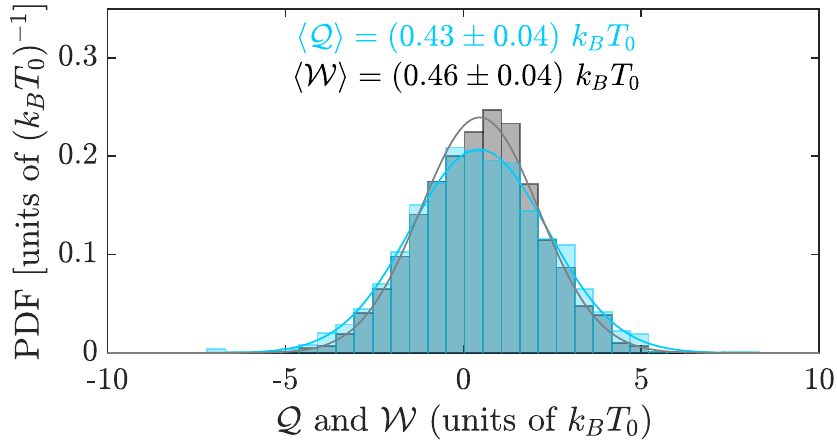}
	\caption{\textbf{Work and heat distribution}. The work (black, right) and heat (light blue, left) mean values and their confidence interval are computed from the $2000$ bit-flip trajectories. The PDF are described by a gaussian shape (fit in continuous line superposed to the histograms).}
	\label{W_Q_bitflip}
\end{figure}

For each of the $2000$ experimental bit-flip trajectories, we compute the value of work and heat to sample their statistics. Their probability distribution function (PDF) are plotted in Fig.~\ref{W_Q_bitflip}: they are both very close to gaussians, with a mean value around $0.45\,k_BT_0$ and a standard deviation around $1.7\,k_BT_0$. We can compute from this data the average work and heat and their uncertainty, as used in the Letter.

Since we access the whole statistics, it could be interesting to probe Fluctuation Theorems for the transformation corresponding to the protocol. We focus here on the work, and since we start from an equilibrium state, we should be entitled to use Jarzynkski's equality~\cite{Jarzynski-1997} which states the out-of-equilibrium work $\W$ is linked to the free energy difference $\Delta F$ between the initial and final state by:
\begin{equation}
\langle e^{-\frac{\W}{k_BT_0}} \rangle = e^{-\frac{\Delta F}{k_BT_0}}
\end{equation}
Specifically, since the initial and final state are the same for a bit-flip, we have $\Delta F=0$ and thus expect $\langle \exp[-\W/(k_BT_0)] \rangle = 1$. However, if we compute this quantity with the experimental data, we get $\langle \exp[-\W/(k_BT_0)] \rangle = 2.5$, equivalent to $\Delta F=0.9 k_BT_0$. 

The purpose of this appendix is to explain the origin of this discrepancy. We show that even a small measurement noise, of variance only 3\% of the thermal noise, is enough to bias strongly the statistics during the exponential averaging of the work in the case of an underdamped dynamics. The fondamental reason for this issue is that both the work and its variance scale as $1/Q$, thus are very small in our underdamped system. The role of the measurement noise is therefore exacerbated when probing Jarzynkski's equality. In the next section, we first study the statistics of the work in the ideal case with no detection error. In the last section, we introduce this measurement uncertainty and demonstrate the strong bias it induces.

\section{Jarzynski's equality in the bitflip protocol}

The work on a single trajectory can be computed using Eq.~(2a) of the Letter:
\begin{align}
\W &= U_0(t_i) - U_1(t_i)+ U_1(t_f) - U_0(t_f),\\
&=\frac{1}{2}k\Big[x(t_i)^2-(|x(t_i)|-X_1)^2\nonumber\\
&\qquad\quad +(|x(t_f)|-X_1)^2-x(t_f)^2\Big],\\
&= kX_1\Big[|x(t_i)|-|x(t_f)|\Big]. \label{eqW3}
\end{align}
Let us assume first that we have no measurement noise, so that Eq.~\ref{eqW3} can be directly used to compute the stochastic work from the measurement. In practice, no trajectory crosses the energetic barrier in the two wells configuration, so that the dynamics is that of an harmonic oscillator evolving in a well whose center is moved during the protocol: from $\pm X_1$ to $0$ at $t=t_i=0$, and from $0$ to $\mp X_1$ at $t=t_f=\mathcal{T}/2$. The system is described by a Langevin equation:
\begin{equation}
m\ddot{x}+\frac{m\omega_0}{Q} \dot{x} + kx = k \mathcal{X}_1(t)+F_\tth(t),
\end{equation}
where $F_\tth(t)$ is the thermal noise driving and $\mathcal{X}_1(t)$ is the deterministic forcing [$\mathcal{X}_1(t)=\pm X_1$ for $t<t_i$, $\mathcal{X}_1(t)=0$ for $t_i<t<t_f$, $\mathcal{X}_1(t)=\mp X_1$ for $t>t_f$]. It is therefore linear can be separated into a deterministic part $x_D(t)$ and a stochastic one $x_\tth(t)$:
\begin{equation}
x(t)=x_D(t)+x_\tth(t),
\end{equation}
where
\begin{align}
m\ddot{x}_D+\frac{m\omega_0}{Q} \dot{x}_D + kx_D &= k \mathcal{X}_1(t),\\
m\ddot{x}_\tth+\frac{m\omega_0}{Q} \dot{x}_\tth + kx_\tth &= F_\tth(t).
\end{align}
The deterministic term $x_D(t)$ is in equilibrium in $U_1$ till $t_i$, evolving in $U_0$ till $t_f$, and back in $U_1$ afterwards: $x_D(t_i)=\pm X_1$ and $x_D(t_f)=\mp (X1-\Delta X)$. The stochastic term $x_\tth(t)$ is simply evolving at all times in a still potential of stiffness $k$ centered in $0$, thus in $U_0'(x)=\frac{1}{2}kx^2$. From Eq.~\ref{eqW3} the work can be further expressed as
\begin{align}
\W &= kX_1\Delta X + kX_1\Big[x_\tth(0)+x_\tth(\mathcal{T}/2)\Big].  \label{eqW4}
\end{align}
Taking the average over many trajectories of the last equation, using $\langle x_\tth \rangle = 0$, we end up with the average stochastic work being equal to the deterministic work $\W_D = k X_1 \Delta X$. The undershoot $\Delta X$ is computed from Eq.~7 of the letter as $\Delta X = \alpha X_1$, with 
\begin{align}
\alpha=1-e^{-\frac{\omega_0\mathcal{T}}{4Q}}\approx \frac{\pi}{2Q},
\end{align}
so that the deterministic work can be written:
\begin{align}
\W_D &= \alpha k X_1^2 \approx \frac{\pi}{2Q}k X_1^2.
\end{align}
In both of these last equations, the approximation stands for $Q\gg 1$ and underline the scaling in $1/Q$ of the deterministic work.

If we want to use Jarzynski's equality to infer the free energy change, the averaging operation is conducted on $\exp[-\W/(k_BT_0)]$, involving this time more than the 1st moment of the work. We therefore need to know the statistics of the stochastic contribution $\Sigma_x = x_\tth(0)+x_\tth(\mathcal{T}/2)$. $\Sigma_x$ is the sum of the position of the oscillator submitted to thermal noise only in the still well $U_0'$ at two instants half a period appart. In absence of dissipation the motion would be purely deterministic and we could write $x_\tth(\mathcal{T}/2)=-x_\tth(0)$, hence $\Sigma_x=0$. For a finite $Q$, this is anyway not true  there is a small undershoot due to dissipation during the oscillation, and the $x_\tth(\mathcal{T}/2)$ lose its correlation with $x_\tth(0)$ due to the thermal noise forcing. For a given trajectory, let us again describe this process by the sum of a deterministic oscillation and a stochastic one, both in the potential $U'_0$:
\begin{equation}
x_\tth(t)=x_D'(t)+x_\tth'(t),
\end{equation}
where
\begin{align}
m\ddot{x}'_D+\frac{m\omega_0}{Q} \dot{x}'_D + kx'_D &= 0,\\
m\ddot{x}'_\tth+\frac{m\omega_0}{Q} \dot{x}'_\tth + kx'_\tth &= F_\tth(t).
\end{align}
The term $x_D'(t)$ describes the deterministic relaxation in the parabolic well $U'_0(x)$ from the stochastic initial condition $x_D'(0)=x_\tth(0)$, $v_D'(0)=v_\tth(0)$, and leads to $x_D'(\mathcal{T}/2)=-(1-\alpha)x_\tth(0)$. The term $x_\tth'$ describes the purely stochastic contribution that starts from $x_\tth'(t_i)=0, v_\tth'(t_i)=0$ and gains some amplitude because of the thermal noise. Eq.~\ref{eqW4} is then simplified to:
\begin{align}
\W &= \W_D +  k X_1 \Big[\alpha x_\tth(0) + x'_\tth(\mathcal{T}/2)\Big].  \label{eqW5}
\end{align}
The statistics of $\W$ thus include two stochastic contributions: one corresponding to the initial position (at time $t_i=0$) of an oscillator at equilibrium in a single well [$x_\tth(0)$]; and one corresponding to the position $x_\tth'$ at $\mathcal{T}/2$ of an oscillator starting at $t=0$ with a zero energy and submitted to thermal noise only. The distribution is thus gaussian, centered on the deterministic work $\W_D$, with a variance
\begin{equation}
\sigma_\W^2 = k^2X_1^2 [\alpha^2 \sigma_x^2 + \sigma_x'^2(t_f)]. \label{sigmaW1}
\end{equation}

To compute $ \sigma_x'^2(t_f)$, we studying the stochastic heat exchanged with the bath~\cite{Dago-2022-PRL}:
\begin{align}
	\langle \dot \Q '\rangle & = \frac{\omega_0}{Q} (2 \langle \K' \rangle - k_B T_0), \label{Qdot6}
\end{align}
where we use primed quantities to designate the harmonic oscillator starting at $t_i=0$ with zero energy. During the evolution in $U'_0$, no work is performed, thus this heat is transferred to the kinetic and potential energy:
\begin{align}
	- \langle \dot \Q' \rangle & = \langle \dot \K' \rangle + \langle \dot U' \rangle = 2 \langle \dot \K' \rangle, \label{Qdot7}
\end{align}
where equipartition is used for the last equality. Combining Eqs.~\ref{Qdot6} and \ref{Qdot7} and integrating this linear first order differential equation from the initial condition $\langle \K'(0) \rangle=0$, we compute
\begin{align}
	\langle U'(\mathcal{T}/2) \rangle & = \langle \K'(\mathcal{T}/2) \rangle = \frac{1}{2}\alpha' k_BT_0,\\
	\mathrm{where}\ \alpha' & = 1-e^{-\frac{\omega_0\mathcal{T}}{2Q}}\approx \frac{\pi}{Q}, \label{eqUtf}
\end{align}
Since $\langle U'(\mathcal{T}/2) \rangle = \frac{1}{2}k\langle x_\tth'^2(\mathcal{T}/2) \rangle = \frac{1}{2}k \sigma_x'^2(\mathcal{T}/2)$, we deduce from Eq.~\ref{sigmaW1} the variance of $\W$:
\begin{align}
\sigma_\W^2 & = k X_1^2 k_BT_0 (\alpha^2+\alpha').
\end{align}
From the expression of $\alpha$ and $\alpha'$, we easily compute $\alpha^2+\alpha'=2\alpha$, hence:
\begin{align}
\sigma_\W^2 & =  2 \W_D k_BT_0. \label{sigmaW}
\end{align}
The variance of $\W$ is proportional to its mean value, thus also scale as $1/Q$. The PDF of $\W$ finally writes:
\begin{align}
P(\W)=\frac{1}{Z_\W} e^{-\frac{(\W-\W_D)^2}{4 \W_D k_BT_0}}.
\end{align}

We can now study Jarzynski's equality:
\begin{align}
\langle e^{-\frac{\W}{k_BT_0}} \rangle & = \int_{-\infty}^\infty d\W P(\W) e^{-\frac{\W}{k_BT_0}}, \\
& =  \int_{-\infty}^\infty d\W \frac{1}{Z_\W} e^{-\frac{(\W+\W_D)^2}{4 \W_D k_BT_0}}, \\
& = 1,
\end{align}
from which we deduce $\Delta F=0$, as theoretically expected since the initial and final potential are the same. We see in this computation that the link between the average and the variance of the work, expressed in Eq.~\ref{sigmaW}, is of utmost importance for Jarzynski's equality to hold.

\section{Influence of the measurement noise}

However, as stated in the first section, from the experimental data we get $\langle \exp[-\W/(k_BT_0)] \rangle = 2.5$, far from the expectation. The reason for this discrepancy is rooted in the measurement noise, which widens the distribution of the measured work, thus breaks the relation between the average and variance of the work of Eq.~\ref{sigmaW}. In our experiment, this noise comes mainly from the shot noise in the interferometer and from high order oscillations modes of the cantilever we use as an oscillator. We can safely assume that this measurement noise is additive, gaussian with a variance $\sigma_n^2$, and uncorrelated on a time $\mathcal{T}/2$. We moreover estimate by studying the power spectrum density of the signal that $\sigma_n^2 \sim 0.03 \sigma_x^2$. From Eq.~\ref{eqW4}, we see that this measurement noise leads to an additional term in $\sigma_\W$:
\begin{equation}
\sigma_\W^2 = 2 \W_D k_BT_0 + 2 k^2X_1^2 \sigma_n^2.
\end{equation}
Let us compare these two terms for $Q=100$ and $X_1=5.3\sigma_x$:
\begin{align}
2 \W_D k_BT_0 & \sim \frac{\pi}{Q} k X_1^2 k_BT_0 \sim 0.9(k_BT_0)^2, \\
2 k^2X_1^2 \sigma_n^2 & \sim 1.7 (k_BT_0)^2.
\end{align}
Though the measurement noise is much smaller than the stochastic signal due to the thermal noise of the oscillator, the former contributes twice as much as the latter in the variance of the work. We note that the estimated total variance in this approach perfectly matches the measured PDF width. We can further compute $\langle \exp[-\W/(k_BT_0)] \rangle$ by using the updated variance of $\W$ in the expression of the PDF, and end up with:
\begin{align}
\langle e^{-\frac{\W}{k_BT_0}} \rangle & = e^{\frac{k^2X_1^2 \sigma_n^2}{k_B^2T_0^2}} \sim 2.3.
\end{align}
This value is again very close to the experimental one, and demonstrates that a variance of the measurement noise at only $3\%$ of the thermal noise is enough to impede the use of Jarzynski's equality to infer $\Delta F$. The fondamental reason for this issue is that both the work and its variance scale as $1/Q$, thus are very small in our underdamped system. We note however that the measurement noise is not an issue to compute the mean value of $\W$, since it averages to 0 for a large number of trajectories.

\section{Data availability}

The data that support the findings of this study are openly available in Zenodo~\cite{Dago-2023-DatasetPRE}.

\bibliography{Bitflip}

\end{document}